\documentclass[12pt, letterpaper]{article}

\usepackage{amssymb}
\usepackage{amsmath}
\usepackage{verbatim}
\usepackage{color}

\mathversion{normal}

\makeatletter \@addtoreset{equation}{section} \makeatother

\newcommand{\bea}{\begin{eqnarray}}
\newcommand{\eea}{\end{eqnarray}}
\newcommand{\be}{\begin{eqnarray}}
\newcommand{\ee}{\end{eqnarray}}
\newcommand{\ben}{\begin{eqnarray*}}
\newcommand{\een}{\end{eqnarray*}}
\newcommand{\beq}{\begin{equation}}
\newcommand{\eeq}{\end{equation}}

\ifx\genfrac\sdflkaj

\else

\fi
\newcommand{\sfrac}[2]{{\textstyle\frac{#1}{#2}}}
\ifx\half\sdflkaj
\newcommand{\half}{\sfrac{1}{2}}
\fi

\let\oldPhi=\Phi
\let\oldPsi=\Psi
\renewcommand{\Phi}{\mathnormal{\oldPhi}}
\renewcommand{\Psi}{\mathnormal{\oldPsi}}

\newenvironment{myeqnarray}{\arraycolsep0pt\begin{eqnarray}}{\end{eqnarray}\ignorespacesafterend}
\newenvironment{myeqnarray*}{\arraycolsep0pt\begin{eqnarray*}}{\end{eqnarray*}\ignorespacesafterend}

\def\[{\begin{equation}}
\def\]{\end{equation}}
\def\<{\begin{myeqnarray}}
\def\>{\end{myeqnarray}}

\ifx\href\asklfhas\newcommand{\href}[2]{#2}\fi

\begin{document}

\begin{titlepage}
\vspace{1in}
\begin{flushright}
\end{flushright}
\vspace{1cm}

\begin{center}
{\Large\bf Comments on the slope function}\\
\vspace{.6in}

{Minkyoo Kim}\footnote{E-mail:
mimkim80@gmail.com}
\vskip 1.5cm
{MTA Lend\" ulet Holographic QFT Group, Wigner Research Centre for physics,
H-1525 Budapest 114, P.O.B. 49, Hungary}\\

\vspace{.6in}
\end{center}
\begin{center}
{\large\bf Abstract}
\end{center}
\begin{center}
\begin{minipage}{5in}

The exact slope function was first proposed in $SL(2)$ sector and generalized to $SU(2)$ sector later.  
In this note, we consider the slope function in $SU(1|1)$ sector of ${\cal N}=4$ SYM. 
We derive the quantity through the method invented by N. Gromov and discuss about its validity.
Further, we give comments on the slope function in deformed SYM.

\end{minipage}
\end{center}
\end{titlepage}

\newpage

\section{Introduction}
Solving interacting QFT is generally very difficult because we still lack general techniques beyond perturbation.
However, if we restrict our interest to a highly symmetric but still interesting model, one can utilize various methods to treat nonperturbative field theory. For instance, the sine(sinh)-Gordon model in $(1+1)$-dimension is a prototypical QFT which could be exactly solved through integrability \cite{Samaj:2013yva}. 
Beyond two dimensional theories, such an example seems to be the maximally supersymmetric, four-dimensional Yang-Mills theory since the theory shows integrable structures in itself and also in holographic dual description.
Integrability structures were found in various observables of $N=4$ SYM such as conformal dimensions of gauge invariant operators, spacetime scattering amplitudes and correlation functions.\footnote{Although there exist many literatures for integrability of $N=4$ SYM, the best exposition would be \cite{Beisert:2010jr} written by leading experts in this area. Very recently, pedagogical lecture notes were written and available in \cite{Bombardelli:2016rwb}.}   
Among those, the spectral problem is the most developed and is now formulated in a beautiful algebraic structure, the so-called quantum spectral curve \cite{Gromov:2014caa}. 
Historically, several key concepts, quantities and equations led to obtain the QSC. For example, the Beiset-Staudacher Bethe ansatz equation \cite{Beisert:2005fw}, exact $S$-matrix \cite{Beisert:2005tm, Arutyunov:2006yd}, $Y,T,Q$-systems \cite{Gromov:2009tv, Bombardelli:2009ns, Arutyunov:2009ur, Gromov:2010km, Gromov:2011cx} are of those. One of such important developments was the exact slope function \cite{Basso:2011rs}. 

In \cite{Basso:2011rs}, Basso gave a conjecture for the slope function of composite operators in $sl(2)$ sector. 
The slope function is defined as a coefficient of the leading anomalous dimension in the small spin $S$ limit such as
\begin{equation}
\Delta-J = S+ \gamma = \alpha_J S +{\cal O}(S^2), \label{slope}
\end{equation}
where $J$ is the number of $Z$ fields consisted of the BPS vacuum and $S$ is the number of the covariant derivatives ${\cal D}$ where $S$ is clearly an integer but can be thought as any real number in a kind of analytical continuation.
Very remarkable facts of the conjecture are its exactness for all-loops of the coupling constant $\lambda$ and that the quantity is independent of wrapping effects. \footnote{On the other hand, the slope function in ABJM model depends on wrapping contributions. Thus, since simple derivation through asympototic BAE or the Baxter equation have been not enough, we needed a formulation beyond asymptotic formalism. Later, the quantum spectral curve allows to calculate the slope function \cite{Gromov:2014eha}. }
After Basso's conjecture, the slope function was formally derived in \cite{Gromov:2012eg, Basso:2012ex} and was also generalized from $SL(2)$ sector to $SU(2)$ sector \cite{Gromov:2012eg}.
Till now, the conjecture have passed some nontrivial tests \cite{Kruczenski:2012aw, Beccaria:2012pm}. Finally, the slope and the curvature which are separately the leading and the next to leading terms of small spin expansion were delicately calculated through the quantum spectral curve \cite{Gromov:2014bva}.

In this letter, we would like to obtain the expression of the slope function in the $SU(1|1)$ sector. 
It will provide a complete table for slope functions in all rank-one sectors. 
Further, it would give another source to check the quantum spectral curve as comparing results in this note. 
We derive the anomalous dimension and consider the small impurity limit which would be a kind of analytic continuation. 
Moreover, we give some comments on the slope function in deformed SYM. 
As the simplest choices, we consider the slope of $SU(2)_{\beta}$ and $SL(2)_{\beta}$  
We conclude this letter with some discussions.


\section{Comments on the slope function} 
The anomalous dimension $\gamma$ in (\ref{slope}) is defined as the second conserved charge ${\cal Q}_2$ such as
\begin{equation}
\gamma = \frac{i \sqrt{\lambda}}{2\pi} \sum_{j=1}^{S} \left(\frac{1}{x_{j}^{+}}-\frac{1}{x_{j}^{-}} \right) \equiv {\cal Q}_{2}
\end{equation}
among infinitely many conserved charges.
The $\gamma$ can be expanded in both of large and small 't Hooft coupling constant $\lambda$ and generally depends on complicated wrapping corrections. Surprisingly, Basso conjectured that it could be computed only from the asymptotic BAE since the slope function of $N=4$ SYM is independent of wrapping effects. 
Such an exact slope is given as
\begin{equation}
\alpha_{J} = 1 + \frac{\Lambda}{J} \frac{I_{J+1}(\Lambda)}{I_{J}(\Lambda)},
\end{equation}
where $\Lambda$ is defined as $n\sqrt{\lambda}$ with the mode number $n$ and the function $I_{n}(x)$ is the modified Bessel function of the first kind. 
After this work, Gromov elegantly derived the slope function by considering a quite special limit \cite{Gromov:2012eg}. 
With this derivation method, we first analyse the slope function in $SU(1|1)$ sector of $N=4$ SYM and discuss about the expression. 
Next, we give comments on the slope function in $\beta$-deformed SYM.

\subsection{Slope function in $SU(1|1)$ sector}
With the following notation,
\begin{eqnarray}
&& x+\frac{1}{x} = \frac{u}{g}, \quad x_{k}^{\pm} = x\left(u_{k} \pm \frac{i}{2} \right) \\
&& x(u) = \frac{u}{2g} + \frac{u}{2g} \sqrt{1- \frac{4g^2}{u^2}} ,
\end{eqnarray}
the  $su(1|1)$ BAE for the composite operator of one type of boson and fermion such as ${\rm tr}(\psi^{J-S} Z^{S})$ can be written as follows\footnote{We used a little different but physically equivalent convention for spectral parameters compared with \cite{Beisert:2005fw}. }  :
\begin{equation}
\left(\frac{x_{k}^{+}}{x_{k}^{-}}\right)^{J} = \prod_{j \neq k}^{S} \frac{1-\frac{1}{x_k^{+} x_j^{-}}}{1-\frac{1}{x_k^{-} x_j^{+}}} \sigma^2 (x_k , x_j), \quad (k=1,\cdots, S) \label{bae}
\end{equation}
Taking the logarithm and dividing by $i$, one gets 
\begin{equation}
\frac{J}{i}\log{\left(\frac{x_{k}^{+}}{x_{k}^{-}}\right)} - \frac{1}{i}\sum_{j \neq k}^{S} \log{\left(\frac{1-\frac{1}{x_k^{+} x_j^{-}}}{1-\frac{1}{x_k^{-} x_j^{+}}} \sigma^2 (x_k , x_j ) \right)} = 2\pi n_{k} . \label{lbae}
\end{equation}
Now we will consider the limit which was used in \cite{Gromov:2012eg} such as
\begin{equation}
n_k = n \rightarrow 0 \quad {\rm for} \,\ \rm{all} \,\ k  , \quad \Lambda = n \sqrt{\lambda} = {\rm fixed}.
\end{equation}
\footnote{We should give an important remark for this limit. The mode numbers $n_{k}$ should usually be different because of the fermionic nature of the $SU(1|1)$ sector. Nevertheless, it was known that roots with different mode numbers do not interact in $n_{k} \rightarrow 0$ limit \cite{Basso:2011rs, Gromov:2012eg}. Thus, one can reintroduce the different mode numbers in the final expression of anomalous dimension.} 
Since this limit is definitely a large $\lambda$ limit, one can make use of strong coupling expressions for Zhukowski variables and the dressing phase such as
\begin{eqnarray}
\log \sigma(x_k ,x_{j}) &\simeq & {\frac{i(x_k -x_{j}) }{g(-1+x_k^2) (-1+x_k x_{j}) (-1+x_{j}^2)} }, \\
x^{\pm}(z) &=& x(z) \pm \frac{i}{2 g} \frac{x^2(z)}{x^2(z) -1} + O(1/g^2), \\
\gamma &\simeq & 2 \sum_{j=1}^{S} \frac{1}{x_j^2 -1} = G(-1) - G(1)
\end{eqnarray}
where we defined the resolvent $G(x)$ such as
\begin{equation}
G(x) = \sum_{j=1}^{S} \frac{1}{x-x_j }. \label{resol}
\end{equation}
With (\ref{lbae}), by multiplying  $\frac{\sqrt{\lambda}(x_k^2 -1)}{4 \pi x_k^2}$ and using the above identities, we get the following form of the BAE :
\begin{equation}
\frac{2J+\gamma}{2x_k} - \sum_{j=1}^{S} \left( \frac{x_j}{x_j^2 -1} \right) \frac{1}{x_k^2} = \frac{\Lambda (x_k^2 -1)}{2 x_k^2} . \label{mbae}
\end{equation}
Next, what we have to do is to multiply $\frac{1}{x- x_k}$ to (\ref{mbae}) and sum over $k$. 
Then, the first term of the l.h.s. of (\ref{mbae}) becomes 
\begin{equation}
\left(\frac{2J+\gamma}{2x} \right) (G(x) - G(0)).
\end{equation}
Also, the second term of the l.h.s. of (\ref{mbae}) can be expressed as
\begin{equation}
\left(\sum_j \frac{x_j }{x_j^2 -1 }\right) \left(\sum_k \frac{1}{x_k^2 } \frac{1}{x-x_k} \right) = -\frac{1}{2} (G(1)+G(-1)) \times \frac{G(x)-G(0) - G'(0)x }{x^2},
\end{equation}
where we used 
\begin{equation}
G(0) = - \sum_k \frac{1}{x_k }, \quad G'(0) = - \sum_k \frac{1}{x_k^2 }.
\end{equation}
Lastly, the r.h.s of (\ref{mbae}) can be also written similarly such as
\begin{equation}
 \sum_{k} \frac{\Lambda(x_k^2 -1)}{2 x_k^2 } \frac{1}{x-x_{k} }  = \frac{\Lambda}{2} G(x) -\frac{\Lambda}{2} \left( \frac{G(x)-G(0) - G'(0)x }{x^2} \right).
\end{equation}
Thus, (\ref{mbae}) can be rewritten such as
\begin{equation}
G(x) = \frac{(G(1)+G(-1)+\Lambda)G(0)+[(G(1)+G(-1)+\Lambda)G'(0)+(2J+\gamma)G(0)]x }{(G(1)+G(-1)+\Lambda) + (2J+\gamma)x - \Lambda x^2}. \label{rbae}
\end{equation}

Now let us consider large $x$ limit of (\ref{rbae}) to replace $G'(0)$ by non-derivative quantities such as $G(1)$, $G(-1)$ and $G(0)$.
In this limit, the resolvent $G(x)$ behaves as $G \sim \frac{S}{x}$ from (\ref{resol}). Thus, by comparing leading contributions from each sides, we get
\begin{equation}
G'(0) = - \frac{\Lambda S + (2J+\gamma)G(0)}{\Lambda + G(1) + G(-1)}. \label{resd}
\end{equation}
Substituting (\ref{resd}) in (\ref{rbae}), we can write the following :
\begin{equation}
G(x) = \frac{(G(1)+G(-1)+\Lambda)G(0)- \Lambda S x }{(G(1)+G(-1)+\Lambda) + (2J+\gamma)x - \Lambda x^2}. \label{resol1}
\end{equation} 
At this stage, we still have some unknown quantities such as $G(1)$, $G(-1)$ and $G(0)$. However, they can be eliminated by checking the consistency of (\ref{resol}) at $x = \pm 1$. They are determined such as
\begin{equation}
G(1)= -\frac{\Lambda  S+\gamma  J}{2J}, \quad G(-1)= \frac{-\Lambda  S+\gamma  J}{2J}, \quad G(0)= -\frac{2 J^3 \gamma +J^2 \gamma^2-S^2 \Lambda^2}{2 J^2 \Lambda -2 J S \Lambda }.
\end{equation}
We finally have
\begin{equation}
G(x) = \frac{\left(-2 J^3 \gamma -J^2 \gamma ^2+S^2 \Lambda ^2\right)-2 \left(J^2 S \Lambda \right) x}{\left(2 J^2 \Lambda -2 J S \Lambda \right)+\left(4 J^3+2 J^2 \gamma \right) x-2 \left(J^2 \Lambda \right) x^2}.
\end{equation}
Note that the resolvent $G(x)$ is analytic. So, if we require vanishing residues at poles of $G(x)$, one can obtain the following possibilities for $\gamma$ :
\begin{equation}
\gamma = -2J - \frac{S \Lambda }{J}, \quad -2J + \frac{S \Lambda }{J}, \quad -S - S \sqrt{1+\frac{\Lambda^2}{J^2}} , \quad -S + S \sqrt{1+\frac{\Lambda^2}{J^2}}.
\end{equation}
The first and second expressions would be inconsistent because they would not become zero in the $S \rightarrow 0$ limit.
Besides, the third one seems unphysical because it means that energy of non-BPS state is lower than that of a BPS.
Therefore, we could take the last expression as a correct expression of the anomalous dimension :
\begin{equation}
\gamma = -S + S \sqrt{1+\frac{\Lambda^2}{J^2}} \rightarrow \sum_{n_{k} =1}^{S} \left(-1 + \sqrt{1+ \frac{\lambda}{J^2}n_{k}^{2}} \right) + {\cal O}(S^2 ),
\end{equation}
where we reintroduced the mode number $n_{k}$. Actually, this dispersion is matched with the result from pp-wave limit of $SU(1|1)$ string \cite{Arutyunov:2005hd}.
Further, as $\gamma$ can be expanded in small and large $\frac{\lambda}{J^2}\equiv\lambda'$ limit, we have 
\begin{eqnarray}
&& \alpha_J =1-\frac{1}{S}\sum_{k=1}^{S} n_{k} + \frac{\lambda}{2 J^2}\frac{1}{S}\sum_{k=1}^{S} n^2_{k}+O(\lambda'^2 ) , \label{slopeftn1} \\
&& \alpha_J =-\frac{1}{S}\sum_{k=1}^{S} n_{k} + \frac{J}{2\sqrt{\lambda}S}\sum_{k=1}^{S}\frac{1}{ n_{k}}+O\left(\frac{1}{\lambda'^{3/2}} \right) , \label{slopeftn2}
\end{eqnarray}
where $\alpha_{J} \equiv 1+ \frac{\gamma}{S}$. 

Moreover, one can get expressions for the higher conserved charges which is defined as
\begin{equation}
{\cal Q}_r = i \frac{\sqrt{\lambda}}{2\pi} \sum_{j=1}^{S} \left[\frac{\left(x_{j}^{+}\right)^{1-r}}{r-1} - \frac{\left(x_{j}^{-}\right)^{1-r}}{r-1} \right] ,
\end{equation}
through their generating function $H(x)$ and $H_{0}(x)$ defined by $n \rightarrow 0$ limit of $H(x)$ \cite{Gromov:2012eg}:
\begin{equation}
H(x) = \sum_{j=1}^{S} \frac{\sqrt{\lambda}}{4\pi i} \log{\left(\frac{1-\frac{x}{x_{j}^{-}}}{1-\frac{x}{x_{j}^{+}}} \right)} = -\frac{1}{2} \sum_{r=1} {\cal Q}_{r+1} x^{r}.
\end{equation}
The $H_{0}(x)$ in $su(1|1)$ is given as
\begin{equation}
H_{0} (x) = \frac{\Lambda^2 S x^2 + \Lambda (J-S)\gamma x}{2 J \Lambda x^2 - 2J(2J+\gamma)x -2\Lambda (J-S)}
\end{equation}
Thus, by expanding $H_{0}(x)$ in polynomials of $x$ with the slope function $\alpha_{J}$, we have the following leading order of conserved charges :
\begin{eqnarray}
{\cal Q}_{3} &=& \frac{(2 J^2 \gamma + J \gamma^2 - S \Lambda^2 )x^2 }{2\Lambda(J-S)}, \label{q3}\\
{\cal Q}_{4} &=&  \frac{(4J^3 \gamma + 4J^2\gamma^2 - 2S\gamma)J x^3 }{2\Lambda^2 (J-S)^2}.\label{q4}
\end{eqnarray} 
Note that this is just a formal expression and we need to substitute $\gamma$ and $\Lambda$ with certain precision into (\ref{q3}) and (\ref{q4})  for an explicit value.

\subsection{Slope function in $SU(2)_{\beta}$}
Before going to the deformed case, let us recall how to generalize from the $SL(2)$ result to the $SU(2)$ one.
In \cite{Gromov:2012eg}, the corresponding expression in $SU(2)$ sector was obtained from the slope function in $SL(2)$ by 
replacing $J$ and $\Lambda$ by $-L$ and $-\Lambda$ with $M=S$ :
\begin{equation}
\alpha_{L} = 1- \frac{\Lambda}{L} \frac{I_{-L+1}(\Lambda)}{I_{-L}(\Lambda)}, \label{su2slope}
\end{equation}  
It was further pointed out that the expression (\ref{su2slope}) is meaningful only for first considering small $\Lambda$ and setting $L$ to integers. 
As the small $\Lambda$ expansion of the modified Bessel function gives the gamma function, one can get the expression (36) of \cite{Gromov:2012eg}. 

For the $SU(2)_{\beta}$ sector of $\beta$-deformed SYM, the asymptotic Bethe equation is given as
\begin{equation}
\left(\frac{x^{+}_{k}}{x^{-}_{k}}\right)^L = \prod_{\stackrel{j=1}{j \neq k}}^{M}
\Bigg\{ 
\left( \frac{x^{+}_{k} - x^{-}_{j}}
{x^{-}_{k} - x^{+}_{j}}\right)
\left(\frac{1- \frac{1}{x^{+}_{k} x^{-}_{j}}}{1- 
\frac{1}{x^{-}_{k} x^{+}_{j}}} \right)
\sigma( u_{k}, u_{j} )^{2} \Bigg\} e^{2\pi i \beta L} \label{su2b}
\end{equation}
where $M$ is the number of magnons in length $L$.
The deformation effect to the BAE is just $e^{2\pi i \beta L}$.  
Therefore, if we repeat similar calculations with (\ref{su2b}), one can use the same expression (\ref{su2slope}) by redefining $\Lambda$ such as $\Lambda \rightarrow \Lambda + {\hat \beta}L$ with ${\hat \beta}\equiv \beta \sqrt{\lambda}$. 
Thus, we simply get the slope function such as
\begin{equation}
\gamma_{\beta} = -M \left( \frac{\Lambda}{L} +{\hat \beta} \right) \frac{I_{-L+1}(\Lambda + {\hat \beta} L)}{I_{-L}(\Lambda + {\hat \beta} L)}.\label{slopebeta}
\end{equation}
Notice that we do not consider any wrapping effects since we just used the asymptotic BAE. Thus, (\ref{slopebeta}) would be valid asymptotically.  

In the regime of $\Lambda'=\Lambda+{\hat \beta} L << 1$, we would have exactly same expression as in undeformed $SU(2)$ :
\begin{equation}
\gamma_{\beta} = \frac{(\Lambda+{\hat \beta} L)^2}{2L(L-1)} - \frac{(\Lambda+{\hat \beta} L)^4 }{8 L (L-1)^2 (L-2)} + \cdots
\end{equation}
If we further expand $\gamma_{\beta}$ in the small ${\hat \beta}$ limit, we can easily express the difference between anomalous dimensions in deformed and undeformed theories as below.
\begin{equation}
\delta\gamma\equiv \gamma_{\beta}-\gamma = M {\hat \beta} \left( \frac{\Lambda }{L-1}-\frac{\Lambda^3 }{2(L-1)^2 (L-2)} \right) + {\cal O}({\hat \beta}^2)
\end{equation}
It still remains to be understood if the expression (\ref{slopebeta}) is valid at large ${\hat \beta}$ regime. Because the generalized slope is delicately defined as we mentioned in the underformed $SU(2)$ sector.

\subsection{Slope function in $SL(2)_{\beta}$}
As the $\beta$-deformation of SYM does not break its conformality and only affect to $SO(6)$ part, we do not have any twisted effects in $SL(2)$ subsector of $SO(4,2)$. However, the quantum string Bethe equation for the TsT-transformations of $AdS_5$ could be studied \cite{McLoughlin:2006cg}. Although the dual gauge theory for this deformation is still unclear, its string theory is well-defined \cite{deLeeuw:2012hp}. 
Here, because the string Bethe equation in $SL(2)_\beta$ sector could be expressed from undeformed $SL(2)$ Bethe equation by considering the shift of the mode number such as
\begin{equation}
n_{k} \rightarrow n_{k} +\beta J, \nonumber
\end{equation}
the fixed parameter $\Lambda$ is also shifted under the mode number shift. Therefore, the slope function for $sl(2)_\beta$ sector can be written as
\begin{equation}
\alpha = 1+ \left(\frac{\Lambda+{\hat \beta}J}{J} \right) \frac{I_{J+1}(\Lambda+{\hat \beta}J)}{I_{J}(\Lambda+{\hat \beta}J)}.
\end{equation}
This can be expanded at the string coupling regime. However, its weak coupling analysis is unclear since it is not obtained from gauge theory.


\subsection{Concluding remarks}
In this letter, we studied the slope function in the $SU(1|1)$ sector of $N=4$ SYM and in integrable twisted models.

We note that we followed the derivation of \cite{Gromov:2012eg} for computation of the slope function in $SU(1|1)$ sector, there exists an alternative derivation method based on Baxter equation as in \cite{Basso:2012ex}. In this derivation, we need to know the long-ranged $su(1|1)$ Baxter equation and its linearized form in the small spin limit.
It would be nice to examine this derivation and compare with our results. 

It would be interesting to derive the results in this letter from string theory.  
The light cone gauged string theory in $SU(1|1)$ have been intensively studied in  \cite{Arutyunov:2005hd}. Also, the short string was exposited via algebraic curve in \cite{Gromov:2011de}. If we can utilize algebraic curve for $SU(1|1)$ sector and consider the short string limit, it may be possible to get the slope function.  
Also, note that the deformation parameter naturally became $\hat{\beta}$ in the slope for $SU(2)_{\beta}$. Interestingly, the metric and various fluxes in Lunin-Maldacena background which is the dual spacetime of $\beta$-deformed SYM are all written in terms of $\hat{\beta}$ \cite{Lunin:2005jy}. So, it would be nice to obtain the slope from direct string computation on  the background with $\hat{\beta}$.

We further remark that we ignored the wrapping effects when we obtained the slope in $SU(2)_{\beta}$ and $SL(2)_{\beta}$. It is known that the slope which is the leading coefficient in small spin limit does not depend on the wrapping in $N=4$ SYM. However, there is no guarantee that such a wrapping independent nature also appears in deformed models. Thus, our results for $SU(2)_{\beta}$ and $SL(2)_{\beta}$ is asymptotically correct. It would be very interesting to check if the wrapping contribution gives any correction. It should be possible since the all-loop formulation is available through the twisted quantum spectral curve \cite{Kazakov:2015efa}.


\section*{Acknowledgements}
We thank Z. Bajnok for valuable comments.
This work was supported by a postdoctoral fellowship of the Hungarian Academy of Sciences, a Lend\"ulet grant and OTKA 116505.

\end{document}